\documentclass[preprint]{revtex4-1}
\usepackage{graphicx}

\def\Title#1{\begin{center} {\Large #1 } \end{center}}
\def\Author#1{\begin{center}{ \sc #1} \end{center}}
\def\Address#1{\begin{center}{ \it #1} \end{center}}

\newcommand\pubblock{\rightline{\begin{tabular}{l} \pubnumber\\
  \pubdate \end{tabular}}}
\newenvironment{Abstract}{\begin{quotation}  }{\end{quotation}}

\newenvironment{Presented}
{\begin{quotation} 
\begin{center} 
 PRESENTED AT\end{center}  
\begin{center}\begin{large}}{\end{large}
\end{center} 
\end{quotation}}

\def\be{\begin{equation}}	\def\ee#1{\label{#1}\end{equation}}
\def\ba{\begin{array}}		\def\ea#1{\label{#1}\end{array}}
\def\eean{\end{eqnarray}}
\def\ba{\begin{array}}		\def\ea{\end{array}}
\def\bea{\begin{eqnarray}}	\def\eea#1{\label{#1}\end{eqnarray}}
\def\eea{\begin{eqnarray}}
\def\eea{\end{eqnarray}}		
\def\ra{\rightarrow}			
\def\N{{\mbox{\scriptsize N}}} \def\H{{\mbox{\scriptsize H}}}
	\def\ci{\cite}	
\def\pa{\partial}	
\def\mc{\mathcal}	\def\GN{\Gamma_\N}
\def\TOT{{\mbox{\scriptsize TOT}}} \def\S{{\mbox{\scriptsize S}}}
 \def\a{{\mbox{\scriptsize a}}}
	
\def\etal{{\it et\ al.}}

  \def\TOT{{_{\rm TOT}}}  
\def\N{{_{\rm N}}}	\def\S{{_{\rm S}}}	

\newcommand\pubnumber{MAXLA-1/19, ``Hydrogen''}
\newcommand\pubdate{\today}

\begin{document}
\begin{titlepage}
\pubblock

\vfill
\Title{The hydrogen atom as relativistic bound system} 
\vspace{5mm}
\Author{Mikhail N. Sergeenko}
\Address{Fr. Skaryna Gomel State University, 104 Sovetskaya St., 
BY-246019, Gomel, Belarus\\
{\rm and}\\
Gomel State Medical University, 5 Lange St., BY-246050, 
Gomel, Belarus\\
{\rm msergeen@usa.com}}
\vfill
\begin{Abstract}
 The hydrogen atom as relativistic bound-state system of a proton and 
an electron in the complex-mass scheme is investigated. 
 Interaction of a proton and an electron in the atom is described 
by the Lorentz-scalar Coulomb potential; the proton structure 
is taken into account. 
 The concept of position dependent particle mass is developed. 
 Relativistic wave equation for two interacting spinless particles 
is derived; asymptotic method is used to solve the equation. 
 Complex eigenmasses for the $H$ atom are obtained. 
 The spin center-of-gravity energy levels for the $H$ atom are 
calculated and compared with ones obtained from solution of some 
known relativistic wave equations 
and tabulated NIST data. 

\vskip 5mm
\noindent Pacs: 11.10.St, 03.65.Ge, 03.65.Pm
\end{Abstract}

\vspace{3mm}
\begin{Presented}
VI-th Congress of physicists of Belarus, ``Congress'2017'', Minsk, 
November 20--23, 2017.
\end{Presented}
\vspace{5mm}
\end{titlepage}

\section{Introduction}\label{intro}
 Hydrogen is the most popular basic element in the Universe. 
It is the simplest atom, comprising only a proton and an electron 
which are stable particles. 
 This simplicity means its properties can be calculated theoretically 
with impressive accuracy. 
 The spherically symmetric Coulomb potential, with interaction strength 
parametrized by dimensionless coupling (``fine structure'') constant, 
$\alpha$, is of particular importance in many realms of physics.  

 In the latest measurements, the Rydberg constant $R$ (the ground-state 
energy), the scaling factor for all transition frequencies in $H$ atom, 
has been determined with a precision approaching 10-11~\ci{NIST_ASD}. 
 This means that all other transitions in the $H$ atom, ranging from 
the microwave to the ultraviolet, can be considered as metrological 
standards to this same level of precision~\ci{MohrTayl,CharlTab93}. 

 Rapid advances in the techniques of laser spectroscopy have shown 
the way for experimental measurements at a comparable level 
of precision. 
 Many highly precise spectroscopic data had been obtained 
experimentally, and these data could now be fully understood in 
terms of atomic structure and quantum numbers. 
 Thus, laboratory and astrophysical spectra could 
be interpreted as specific classified transitions between energy 
levels of atoms~\ci{MohrTayl}. 

 At the present time, there is no strict theory of the $H$ atom 
as relativistic two-body system. 
 An outline of the problems encountered in the theoretical calculations 
and explanations where the current uncertainties lie have been reviewed 
in~\ci{CagnEtal}. 
 The description of bound states in a way fully consistent with all 
requirements imposed by special relativity and within the framework 
of relativistic Quantum Field Theory is one of the great challenges 
in theoretical elementary particle physics. 
 Comprehensive description of the $H$ atom is reduced to relativistic 
bound state problem. 

 In this work 
we consider the $H$ atom as relativistic two-body system in 
the potential approach; here we have two fundamental problems: 
1)~two-particle relativistic equation of motion and 
2)~absence of a strict definition of the potential in relativistic 
theory. 
 There are various relativistic effects such as fine and hyperfine 
splitting of the energy levels that should be taken into account 
in more accurate analysis. 
 An objective of this work is to calculate the spin center-of-gravity 
energy levels for the $H$ atom in the framework of developed approach 
and compare them with ones obtained from solution of the Klein-Gordon 
(KG), spinless Salpeter (SS) equations and tabulated NIST data. 

 The Coulomb potential is treated as the Lorentz-scalar function 
of the spatial variable $r$ additive to particle's masses. 
 We develop the concept of distance-dependent particle mass and  
derive a two-particle equation of motion. 
 We use the fact that in relativistic kinematics the spatial 
two-particle relative momentum is relativistic invariant, 
spatial part of Minkowski force are relativistic invariants. 
 We derive relativistic two-particle wave equation and give its 
analytic solution in the form of a standing wave. 
 The free particle hypothesis for the bound state is developed: 
proton and electron move as free particles inside atom on the ends 
of a 3D string~\ci{MyPRA96}. 

\section{Nonrelativistic approximation and relativistic\\ 
 corrections}\label{RelCorrEqs}
 Description of the $H$ atom in nonrelativistic formulation 
is very accurate by engineering standards; however, it is not exact. 
 The usual approach is to take the nonrelativistic approximation as 
the starting point. 
 Then corrections are applied using perturbation theory.

 The energy levels of $H$-like atoms are determined mainly by 
the Dirac eigenvalue, Quantum ElectroDynamic effects such as self 
energy and vacuum polarization, nuclear size and motion effects. 
 The binding energy of an electron in a static Coulomb field 
(the external electric field of a point nucleus of charge $Ze$ with 
infinite mass) is determined predominantly by the Dirac eigenvalue. 
 A brief summary of the theory of the energy levels of the $H$ atom 
relevant to the determination of the Rydberg constant $R$ based on 
measurements of transition frequencies is given 
in~\ci{MohrTayl,CharlTab93}. 

 In nonrelativistic formulation, the $H$ atom is described by 
the Schr\"odinger equation and is usually considered as 
an electron moving in the external field generated by the proton 
static electric field given by the Coulomb potential, 
\be 
V(r)=-\frac\alpha r,\quad\quad\quad 
\alpha=\frac{e^2}{4\pi\epsilon_0\hbar c}\simeq \frac 1{137}. 
\ee{CoulPot}
 The fine structure constant, $\alpha$, combines the constants 
$e^2/4\pi\epsilon_0$ from electromagnetism, $\hbar$ from quantum 
mechanics, and the speed of light $c$ from relativity into 
the dimensional number, which one of the most important numbers 
in all of physics. 

 Solution of the Schr\"odinger's equation for the potential 
(\ref{CoulPot}) gives, for the energy eigenvalues, 
\be 
E_\N=\frac{m_e}2 {\rm v}_\N^2,\quad\quad\quad {\rm v}_\N
=\frac{i\alpha c}N,
\ee{SchreEn} 
where $N=k+l+1$ is principal quantum number, $k$ and $l$ being 
the radial and orbit quantum numbers. 
 The {\it total} energy eigenvalues (\ref{SchreEn}) has the form 
of the kinetic energy for a {\it free} particle with the imaginary 
discrete velocity, $v_\N=i\alpha c/N$. 
 This means, that the motion of the electron in a hydrogen atom 
is free, but restricted by the ``walls'' of 
the potential~\ci{MyEPJC12,MyPRA96}. 

 The Rydberg constant, 
\be 
R_\infty=\frac{m_e(\alpha c)^2}2 \frac 1{\hbar c}
=10\,973\,731.568\,549(83)\  m^{-1},
\ee{Rydb}
is defined in terms of a free electron mass, $m_e$. 
 The dimension of the Rydberg is inverse distance, so it is directly 
related to wave number (inverse of wavelength) measurements. 
 The ionization energy of the hydrogen ground state (the Rydberg energy),  
\be 
E_0=\frac{m_e(\alpha c)^2}2\equiv R_\infty\hbar c
=13.598\,434\,485\,644(12)\,eV,
\ee{RydbEner}
is not exactly 1 Rydberg multiplied by $\hbar c$. 
 In nonrelativistic formulation, the $H$ atom is modeled as a particle 
of reduced mass~$\mu=m_e m_p/(m_e+m_p)$, which is used for calculations 
in (\ref{Rydb}) instead of $m_e$ as a correction. 

 According to the expression (\ref{SchreEn}), all energy eigenfunctions 
$\psi_{nlm}$ with the same value of $n$ have the same energy $E_n$ and 
they should show up as a single line in an experimental line spectrum;
this is the spin center-of-gravity energy level. 
 When these spectra are examined very precisely, the energy levels for 
a given value of $n$ are found to consist of several closely spaced 
lines, rather than a single one; that is the $H$ atom fine structure. 

 There are three types of relativistic corrections in the $H$ atom 
energy levels which can be listed in order of decreasing magnitude: 
$\alpha)$ fine structure (includes three effects: Einstein's 
relativistic correction, spin-orbit interaction, the Darwin term for 
states of zero angular momentum);
$\beta)$ Lamb shift (includes virtual photons and virtual 
$e^+e^-$ pairs); 
$\gamma)$ hyperfine splitting (couples the spins of proton and electron). 

 The spectroscopic data are usually analyzed with the use of 
the Sommerfeld's fine-structure formula~\ci{Bohm79}, 
\be 
E_{nj}=m_e c^2\Biggl\{1+\left[\frac{Z\alpha}
{n-(j+\frac 12)+\lambda(j)}\right]^2\Biggr\}^{-1/2},
\ee{SommFinStr}
where $\lambda(j)=\sqrt{\left(j+\frac 12\right)^2-(Z\alpha)^2}$, 
$n=1,\,2,\,\dots$ is principal and $j$ is total quantum numbers. 
 We are interested in the spectrum of photons emitted by a $H$ atom 
when an electron transfers from one energy level to another. 
 This means that the rest mass of the electron is not of concern. 
One defines the observed energy levels of the $H$ atom to be: 
\be 
\mc{E}_{nj}=E_{nj}-m_ec^2.
\ee{TnKin}

 Note the following in (\ref{SommFinStr}). 
 The term $\lambda(j)$ becomes complex if $Z\alpha>j+\frac 12$. 
This means that the $S$ states start to be destroyed above $Z=137$, 
and that the $P$ states begin being destroyed above $Z=274$. 
 Note that this differs from the result of the KG equation, 
which predicts $S$ states being destroyed above $Z=68$ and $P$ states 
destroyed above $Z=82$. 
 Besides, the radial $S$-wave function $R(r)$ diverges as $r\ra 0$, 
i.\,e., $R(r)\sim r^\beta$, with 
$\beta=-\frac 12+\lambda(0)$. 
What is the reason of the problems?  

 The formula (\ref{SommFinStr}), as well as the one predicted from 
solution of the KG wave equation, obtained for 
the Lorentz-vector Coulomb potential. 
 In general there are two different relativistic versions: 
the potential is considered either as the zero component of 
a four-vector, or as a Lorentz-scalar~\ci{SahuAll89,Rajak95}.  
 The relativistic correction for the case of the Lorentz-vector 
potential is different from that for the case of the Lorentz-scalar 
potential~\ci{MyMPLA97}.

\section{The $H$ atom potential}\label{AtomHpot}
 The potential is correctly defined in nonrelativistic theory, 
that gives reasonable results for the $H$ atom spectra. 
 The use of the same nonrelativistic potential (\ref{CoulPot}) 
in relativistic kinematics gives even more accurate results for 
the $H$ atom spectra. 
 The non-relativistic potential model has proven extremely successful 
for the description not only of nonrelativistic systems but also 
relativistic (bound) states. 
 This success is somewhat puzzling in that it persists even when 
the model is applied to relativistic systems like hadrons. 
 Potential models work much better than one would naively 
expect~\ci{LuchSho12}. 

 There is no a rigorous definition of the potential in relativistic 
theory. 
 Relativistically, the electron and proton are affected by virtual 
photons and virtual electron-positron pairs. 
 For example, derivation of the Darwin term does succeed in its 
explaining fully within the nonrelativistic picture alone. 
 First assumption is that the electric potential of the proton does 
not really become infinite as $1/r$ at $r\ra 0$, but is smoothed out 
over some finite proton size. 
 Secondly, the electron ``does not see'' this potential sharply, but 
at a smallest typical distance equal to the Compton wave length 
$\hbar/m_ec$; its uncertainty in position is of the order of 
the Compton wave length. 
 The electron just cannot figure out where the right potential is 
at $r\ra 0$. 

 An electron in physics is considered to be structureless point-like 
particle, but a proton has a complicate structure characterized by 
its form-factor; this structure should be taken into account in
accurate calculations. 
 This means that in the fine structure constant, 
$\alpha=e^2/(4\pi\epsilon_0\hbar c)$, where $e^2=q_e q_p$,   
the proton charge $q_p$ should be replaced by the proton form-factor, 
$eF_E^p(Q)$. 
 We use the proton electric form-factor which is very well described 
[up to $Q^2\simeq 30\,(GeV/c)^2]$ by the dipole formula: 
$F_E^p(Q)=\left[\Lambda^2/(Q^2+\Lambda^2)\right]^2$ where 
$\Lambda\approx 0.84\,GeV/c$~\ci{Stoler93}. 
  This results in the modification of the fine structure constant 
for the $H$ atom in momentum space (hereafter $\hbar=c=1$),  
\be
\alpha_\H^{~}(Q) =\alpha\left(\frac{\Lambda^2}{Q^2+\Lambda^2}\right)^2.
\ee{VHyQ} 
 What is important in this modification? 
 We are interesting in asymptotic properties of $\alpha_\H^{~}(Q)$, 
which are: $\alpha_\H^{~}(Q)\ra\alpha$ at $Q\ra 0$ ($r\ra\infty$) 
and $\alpha_\H^{~}(Q)\ra 0$ at $Q\ra\infty$ ($r\ra 0$). 
 Thus, using the mnemonic rule $Q\ra 1/r$~\ci{MyEPJC12}, we come 
to the following ansatz, for $\alpha_\H^{~}(r)$ in configuration space: 
\be
\alpha_\H^{~}(r)=
\alpha\left(\frac{\Lambda^2 r^2}{1+\Lambda^2 r^2}\right)^2. 
\ee{alfHr}

 Therefore, the Coulomb potential (for the $H$ atom) is modified 
as follows: 
\be
V_\H(r) =-\frac{\alpha_\H^{~}(r)}r.
\ee{VHr}
 The form of the proton form-factor is not important in our 
asymptotic approach; important is its asymptotic behavior. 
 The running coupling $\alpha_\H^{~}(r)$ has the properties: 
$\alpha_\H^{~}(r\ra 0)\ra 0$ and $\alpha_\H^{~}(r\ra\infty)\ra\alpha$. 
 We see similarity with the running coupling in QCD: 
$\alpha_\S(r)$ is frozen at $r\ra\infty$ and is in agreement 
with the asymptotic freedom properties [$\alpha_\S(r\ra 0)\ra 0$]. 

 In case of a bound state, we have a closed system, no external 
field and any particle of the system can be considered as moving 
source of the interaction field. 
 In this case the interacting particles and the potential is 
a~unified system. 

 A serious problem of relativistic potential models is definition 
and the nature of the potential: whether it is Lorentz-vector or 
Lorentz-scalar or their mixture? 
 This problem is very important, for example, in hadron physics where, 
for the vector-like confining potential, there are no normalizable 
solutions; there are normalizable solutions for scalarlike 
potentials~\ci{Sucher95,SemayCeu93}.  
 This issue was investigated in~\ci{Huang01,MyZPhC94}; it was shown 
that the effective interaction has to be Lorentz-scalar in order 
to confine quarks and gluons. 

 A Schr\"odinger-like relativistic wave equation of motion for 
the Lorentz-scalar potential was formulated in~\ci{Huang01}. 
 Though the physical meaning of the Lorentz-scalar potential is not
well described, it is generally accepted that the Lorentz-scalar 
potential is coupled to the rest mass of the particle such that 
the relativistic energy-momentum relation is given 
as~\ci{SahuAll89,Rajak95} 
\be 
E^2-\mathbf{p}^2=\left[m+S(\mathbf{r})\right]^2.
\ee{ScalVrEq}
 This relation is a consequence of the Lagrange equations of 
relativistic motion, with a scaled time as the evolution 
parameter~\ci{Huang01}. 

 The Lorentz-scalar potentials result in concept of the 
position-dependent masses, as usually investigated; 
one assumes that the Lorentz-scalar potential $S(\mathbf{r})$ 
depends only on position \textbf{r}. 
 Application of this Schr\"odinger-like formalism for 
the Lorentz-scalar square-step potential was shown to be 
free from the Klein's paradox~\ci{Huang01}. 
 The predictions are free from not only the Klein's paradox, 
but also the paradoxical results predicted for the Lorentz-vector 
potential~\ci{Huang01}. 

 For the Coulomb potential, this formalism yields the exact 
bound-state  eigenfunctions and eigenenergies~\ci{Huang01},
\be 
E_{nl}=mc^2\Biggl\{1-\left[\frac{Z\alpha}
{n-(l+\frac 12)+\lambda'(l)}\right]^2\Biggr\}^{1/2},
\ee{ScalPotEn}
where $\lambda'(l)=\sqrt{\left(l+\frac 12\right)^2+(Z\alpha)^2}$. 
 These eigenenergies are the same as those predicted exactly from 
solution of the relativistic semi-classical wave 
equation~\ci{MyPRA96,MyMPLA97}. 
 Comparisons between the cases of the Lorentz-scalar potential and 
the Lorentz-vector potential was given in~\ci{MyMPLA97,Huang01}. 
 In contrast to the Lorentz-vector potential, the radial $S$-wave 
function $R(r)$ for the Lorentz-scalar one is regular as $r\ra 0$, 
i.\,e., $R(r)\sim r^\beta$ with 
$\beta=-\frac 12+\lambda'(0)$~\ci{Huang01}. 

\section{Relativistic bound-state equation}\label{Rel2B}
 There are several forms of relativistic two-particle wave equations 
such as the KG equation, the Dirac equation, quasipotential 
equations. 
 The appropriate tool to the description of bound states is QFT. 
 Within the framework of QFT the adequate Poincar\'e-covariant 
description of bound states is the Bethe-Salpeter (BS) 
formalism~\ci{Nakan69,SalBeth51,GelMaLow51}. 
 Description of the bound system is founded on the four-dimensional 
covariant BS equation~\ci{SalBeth51}. 
 The homogeneous BS equation governs all the bound states. 

 However, attempts to apply the BS formalism to relativistic 
bound-state problems give series of difficulties~\ci{LuchSho99}. 
 Its inherent complexity usually prevents to find the exact 
solutions or results in the appearance of excitations in the relative 
time variable of the bound-state constituents (abnormal solutions), 
which are difficult to interpret in the framework of quantum physics. 
 The relative time $t$ in BS formalism causes troubles because the~BS 
wave functions with time-like relative coordinate have no physical 
meaning. 
 The BS equation in the ladder approximation possesses negative norm 
or ghost states, due to its treatment of the relative time degree 
of freedom --- spoiling the naive interpretation of it as a~quantum 
wave equation. 

 For various practical reasons and applications to both QED and QCD 
simplified equations, situated along a path of nonrelativistic 
reduction, are used. 
 More valuable are methods which provide either exact or approximate 
analytic solutions for various forms of differential equations. 
 They may be remedied in three-dimensional reductions of the BS 
equation; the most well-known of the resulting bound-state equations 
is the one proposed by Salpeter~\ci{Salpet52}.

 The most straightforward way out is the reduction of the BS 
equation to the Salpeter equation by a series of approximations, 
which rely on the simplifying assumptions~\ci{LuchSho99}.  
 In order to avoid the problem of relative time, the constant $t$ 
formalism can be used. 
 When one of the constituents is infinitely heavy, the solution with 
the retarded propagator gives the correct results. 
 After applying all these assumptions and approximations 
to the BS equation, one comes to the SS Schr\"odinger-like 
relativistic wave equation~\ci{LuchSho99}, 
$\hat H\psi=\sf{M}\psi$, with the Hamiltonian in the c.m. rest 
frame, 
\be 
\hat H=\sqrt{\hat\mathbf{p}^2+m_1^2}+\sqrt{\hat\mathbf{p}^2+m_2^2}
+\mathsf{W}(\mathbf{r}),
\ee{SalpEq}
where \textbf{p}=\textbf{p}$_1$=-\textbf{p}$_2$. 
 The potential $\mathsf{W}(\mathbf{r})$ in~(\ref{SalpEq}) arises 
as the Fourier transform of the BS kernel $\mc{K}(\mathbf{q})$. 

 The causal condition leads to the BS equation that reduces to 
the SS equation (\ref{SalpEq}) for an instantaneous interaction. 
 The SS equation~(\ref{SalpEq}) is the conceptually simplest 
bound-state wave equation incorporating to some extent relativistic 
effects. 
 This equation has to be regarded as a well-defined standard 
approximation to the BS formalism. 
 When applied, to the $H$ atom, the SS equation contains a projection 
operator onto positive energy states and is different from 
the Dirac equation. 

 However, even this simplest SS equation (\ref{SalpEq}) leads 
to difficulties. 
 The square root of the operators cannot be used as it stands; 
it would have to be expanded in a power series before the momentum 
operator, raised to a power in each term, could act on~$\psi$. 
 There is another problem: the $S$-wave solution of the radial 
equation for the Coulomb potential diverges at the spatial origin 
and behaves as $\psi\sim r^{-\alpha/\pi}$ at $r\ra 0$~\ci{Durand85}.
 This divergence at the spatial origin is actually a general problem 
affecting relativistic wave equations with the Lorentz-vector potential. 
 For example, the solution of the Dirac equation with the Coulomb 
potential for the $S$-wave states behaves at $r\ra 0$ as 
$\psi(r)\sim r^\beta$, $\beta={\sqrt{1-\alpha^2}-1}$. 

\subsection{Transformation of the $SS$ equation}\label{TransSS}
The corresponding to (\ref{SalpEq}) integral of motion (invariant 
bound-state mass) is 
\be
\mathsf{M}=\sqrt{\mathbf{p}^2+m_1^2}+\sqrt{\mathbf{p}^2+m_2^2}
+\mathsf{W}(\mathbf{r}).
\ee{TwoCMDyn}
 Relativistic total energy of a particle $i$, 
$\epsilon_i(\mathbf{p})=\sqrt{\mathbf{p}^2+m_i^2}$ ($i=1,\,2$), 
can be represented as sum of the  kinetic energy, 
$\tau_i(\mathbf{p})$, and the rest mass, $m_i$, i.\,e., 
$\epsilon_i(\mathbf{p})=\tau_i(\mathbf{p})+m_i$. 
 Kinetic and potential energies are different types of the total 
energy. 
 The potential $\mathsf{W}(\mathbf{r})$ is Lorentz-scalar 
as well as particle masses. 
 This is why, it is natural to combine the scalar potential 
$\mathsf{W}(\mathbf{r})$ with masses of particles. 

 So, (\ref{TwoCMDyn}) can be rewritten in the form, 
(see also~\ci {Huang01,SahuAll89,Rajak95}), 
\be 
\mathsf{M}
=\left[\tau_1(\mathbf{p})+w_1(\mathbf{r})\right]
+\left[\tau_2(\mathbf{p})+w_2(\mathbf{r})\right],
\ee{TwoKinMasV}
where we have introduced the position-dependent particle masses  
$w_i(\mathbf{r})=m_i+\frac 12 \mathsf{W}(\mathbf{r})$~\ci{Ono82,Ravn82}. 
 The weight coefficient $\frac 12$ is a subject of discussion.  
 We choose $\frac 12$ because forces acting on bound particles 
in the c.m. rest frame are equal and opposite each other. 
 The Minkowsky force acting on proton and electron is expressed 
via the potential~$\mathsf{W}(\mathbf{r})$. 
 Another ground to the weight coefficient $\frac 12$ gives 
the siring theory~\ci{SonnWeis15,SonnWeis14}. 

 Thus, the total energies of the interacting particles can be 
written as 
$\epsilon_i(\mathbf{p})=\sqrt{\mathbf{p}^2+w_i^2(\mathbf{r})}$ 
in agreement with (\ref{ScalVrEq}) as shown in~\ci{SahuAll89,Rajak95}, 
and the classic dynamical equation for the total invariant 
mass~(\ref{TwoKinMasV}) (the system's total energy in the c.m. 
rest frame) takes the form: 
\be
\mathsf{M}=\sqrt{\mathbf{p}^2+w_1^2(\mathbf{r})}
+\sqrt{\mathbf{p}^2+w_2^2(\mathbf{r})}.
\ee{TwoCMmr}
 This expression can be transformed to the~squared relative momentum, 
\be
\mathbf{p}^2=\frac 1{4\mathsf{M}^2}\left[\mathsf{M}^2-m_-^2\right]
\left[\mathsf{M}^2-(m_+ +\mathsf{W})^2\right], 
\ee{Invp2}
where $m_+=m_1+m_2$, $m_-=m_1-m_2$. 

 The equation (\ref{Invp2}) with the~help of the fundamental 
correspondence principle (according to which physical quantities 
are replaced by operators acting onto the wave 
function)~\ci{MyPRA96,MyMPLA97} results in the two-particle 
spinless wave equation, 
\be 
\Bigl\{\vec{\nabla}^2+K(\mathsf{M}^2)
\left[\mathsf{M}^2-\mc{M}^2(\mathbf{r})\right]\Bigr\}\psi(\mathbf{r})=0,
\ee{Rel2Eq}
where $K(\mathsf{M}^2)=(1-m_-^2/\mathsf{M}^2)/4$, $\mc{M}(\mathbf{r})
= m_++\mathsf{W}(\mathbf{r})$. 
 This equation coincides with the one obtained from the two-body
Dirac Hamiltonian in the rest frame~\ci{SemayCeu93},
\be
H=(\vec\alpha_1-\vec\alpha_2)\vec p +\beta_1m_1+\beta_2m_2
+\frac 12(\beta_1+\beta_2)V(r), 
\ee{DirH2} 
for the scalar potential (note here the weight multiplier $1/2$, 
which has been used for derivation of (\ref{Rel2Eq})). 
 Wave equation (\ref{Rel2Eq}) describes the system of two relativistic 
spinless particles interacting by Lorentz-scalar central potential.

\section{Solution of the QC wave equation}\label{QCsol}
 It is a~problem to solve (\ref{Rel2Eq}) by known methods 
for the~potential~(\ref{VHr}). 
 Instead, we use the asymptotic quasiclassical (QC) method which 
is the mathematical realization of the correspondence principle. 
 The QC method developed in~\ci{MyMPLA97,MyPRA96,MyMPLA00,MyFluc98} 
was tested as the exact for {\em all} one-particle solvable 
spherically symmetric potentials. 
 The corresponding eigenfunctions have the same behavior as the 
asymptotes of the exact solutions~\ci{MyPRA96}. 

 Derivation of the~QC equation in our case is reduced to replacement 
of the operator $\vec\nabla^2$ by the~canonical operator 
$\Delta^c$~\ci{MyPRA96} without the~first derivatives, acting onto 
the~state function   
\be
\Psi(\mathbf{r})=\sqrt{\det\,g_{ij}}\psi(\mathbf{r}), 
\ee{Psi}
where $g_{ij}$ is the metric tensor. 

 In particular case of the spherical coordinates 
$q=\{r,\,\theta,\,\varphi\}$ (the determinant of metric tensor, 
$\det\,g_{ij}=r^2\sin\,\theta$), the canonical operator is~\ci{MyPRA96} 
\be 
\Delta^c=\frac{\pa ^2}{\pa r^2}
+\frac 1{r^2}\frac{\pa^2}{\pa\theta^2}
+\frac 1{r^2\sin^2\theta}\frac{\pa^2}{\pa\varphi^2},
\ee{CanOper}
and the relation (\ref{Psi}) follows from the identity for 
the normalization condition, 
$$
\int\left|\Psi(\mathbf{r})\right|^2dr\,d\theta\,d\varphi 
\equiv\int\left|\psi(\mathbf{r})\right|^2\det\,g_{ij}
dr\,d\theta\,d\varphi= 1. 
$$
 The generalized two-particle QC wave equation in arbitrary 
coordinates has the form
\be 
\Biggl\{\sum_{i=1}^3\left(\frac\pa{g_{ii}\pa q_i}\right)^2
+K(\mathsf{M}^2)\left[\mathsf{M}^2-\mc{M}^2(\mathbf{r})
\right]\Biggr\}\Psi(\mathbf{r})=0.
\ee{Gen2QC}
 The QC wave equation (\ref{Gen2QC}) is the second-order 
differential equation of the Schr\"odinger's type 
in canonical form. 

 Thus, instead of (\ref{Rel2Eq}) we solve the QC equation, 
which is, for the spherically symmetric potentials, 
\be 
\Biggl\{\Delta^c + 
K(\mathsf{M}^2)\biggl[\mathsf{M}^2-\mc{M}(r)^2\biggr]\Biggr\}
\Psi(\mathbf{r})=0. 
\ee{Rel2Equa}
 One of important feature of this equation is that, for two and 
more turning-point problems, it can be solved exactly 
(for eigenvalues) by the conventional leading order in $\hbar$ 
WKB method~\ci{MyPRA96,MyMPLA00}.

 The appropriate solution method of the QC wave equation, which is 
the same for nonrelativistic and relativistic systems, was developed 
in~\cite{MyPRA96,MyMPLA00}.  
 In this method, each of the one-dimensional equations obtained 
after separation of the QC wave equation is solved by the same 
QC method. 
 Equation (\ref{Rel2Equa}) for the~spherical potential~(\ref{VHr}) 
is separated that gives the radial [$s=\mathsf{M}^2$, 
$\mc{M}_\H(r)=m_p+m_e-\alpha_\H^{~}(r)/r$],
\be 
\Biggl\{\frac{d^2}{dr^2}
+K(s)\biggl[s-\mc{M}_\H^2(r)\biggr] 
-\frac{\vec{M}^2}{r^2}\Biggr\}\textrm{R}(r)=0,
\ee{QCrad}
and the~angular, 
\be
\left[\frac{\pa^2}{\pa\theta^2}+\frac 1{\sin^2\theta}
\frac{\pa^2}{\pa\varphi^2} + \vec{M}^2\right]
\sf{Y}(\theta,\varphi)=0,
\ee{AngQC} 
equations. 

 Solution of the~angular equation (\ref{AngQC}) was obtained 
in~\ci{MyPRA96} by the~QC method in the~complex plane, that gives 
$|\vec{M}|=\mathrm{M}_l=\left(l+\frac 12\right)\hbar$, 
for the~angular momentum eigenvalues. 
 These angular eigenmomenta are universal for all spherically symmetric 
potentials in nonrelativistic and relativistic cases~\ci{MyMPLA00}. 

 The radial equation (\ref{QCrad}) cannot be solved analytically by 
standard methods. 
 It can be solved by the same QC method in the complex plane. 
 The equation (\ref{QCrad}) has two turning points and QC quantization 
condition is
\be
\oint_C\sqrt{K(s)\biggl[s-\mc{M}_\H^2(r)\biggr]-\frac{\mathrm{M}_l^2}
{r^2}}dr = 2\pi\left(k+\frac 12\right). 
\ee{FasIc}

 To calculate the phase-space integral (\ref{FasIc}) in the complex 
plane we chose a contour $C$ enclosing the cut between turning points 
$r_1$, $r_2$ at $r>0$, and zeros of the radial state function 
$\textrm{R}(r)$. 
 Outside the contour $C$, the problem has two singularities, i.e., 
at $r=0$ and $\infty$. 
 Using the standard method of stereographic projection and residue 
theory, we should exclude the singularities outside the contour 
$C$~\ci{MyPRA96}. 
 Excluding these singularities we have, for the integral (\ref{FasIc}), 
$I=I_0+I_\infty$, where $I_0=-2\pi|l+\frac 12|$ is contribution of 
the centrifugal term. 
 The phase-space integral $I_{\infty}$ is calculated with the help 
of the replacement of variable, $z=1/r$. 
 Here we have took into account the asymptotic properties of 
the fine coupling (\ref{alfHr}) and its derivative: 
$\alpha_\H^{~}(r)=0$, $\alpha_\H^\prime(r)=0$ at 
$r\ra 0$~\ci{MyEPJC12,MyNDA17}. 
 The integration result is 
\be
I_\infty =\pi\alpha\,m_+\sqrt{\frac{s-m_-^2}{s(-s+m_+^2)}},
\ee{FasIntC}
where $m_+=m_p+m_e$, $m_-=m_p-m_e$. 
 Combining the integration results and condition (\ref{FasIc}) 
we obtain the square equation, 
\be
s^2-4{\rm e}_\N^2s -(2m_\a m_- v_\N)^2=0,
\ee{SquaEq} 
where ${\rm e}_\N^2 = m_\a^2\left(1-v_\N^2\right)$, $m_\a=m_+/2$, 
$v_\N^{~}=\frac 12\alpha/N$, 
and the principal quantum number in our asymptotic approach is 
$N=\left(k+\frac 12\right)+|l+\frac 12|$. 
 Solution of (\ref{SquaEq}) gives, for the~squared eigenmasses 
of the~$H$ atom, 
\be 
s_\N^{~}\equiv \mathsf{M}_\N^2
=2{\rm e}_\N^2 \pm 2\sqrt{\left({\rm e}_\N^2\right)^2
+(m_\a m_-v_\N^{~})^2}. 
\ee{M2Coul}
 Using the~identity in (\ref{M2Coul}), 
\begin{center} 
$\left({\rm e}_\N^2\right)^2+(m_\a m_-v_\N^{~})^2\equiv  
\epsilon_\N^2\left(\epsilon_\N^2\right)^*=|\epsilon_\N^2|^2$,\\
$\epsilon_\N^2 = {\rm e}_\N^2 +im_\a m_-v_\N^{~}$, \ \ \ 
$\left(\epsilon_\N^2\right)^*={\rm e}_\N^2 -im_\a m_-v_\N^{~}$, 
\end{center} 
we have 
\bea
\mathsf{M}_\N^2 =\left(\sqrt{\epsilon_\N^2}
 \pm\sqrt{(\epsilon_\N^2)^*}\right)^2, \label{M2CouCx}\\
 \mathsf{M}_\N = \pm\left(\epsilon_\N^{~}\pm \epsilon_\N^*\right), 
 \label{MCouCx}\\
\epsilon_\N^{~}=\pm\sqrt{\epsilon_\N^2}
=\pm\left({\rm Re}\{\epsilon_\N^{~}\} 
 + i\xi{\rm Im}\{\epsilon_\N^{~}\}\right), \label{wN}\\
{\rm Re}\{\epsilon_\N^{~}\} = \sqrt{\left(|\epsilon_\N^2|
 +{\rm Re}\{\epsilon_\N^2\}\right)/2}, \label{RewN}\\ 
{\rm Im}\{\epsilon_\N^{~}\} = \sqrt{\left(|\epsilon_\N^2|
 -{\rm Re}\{\epsilon_\N^2\}\right)/2}; \label{ImwN}  
\eea
 here in (\ref{wN}) $\xi={\rm sgn}({\rm Im}\{\epsilon_\N^2\})$. 

 Thus, the $H$-atom's eigenmasses are complex. 
 Simple expression (\ref{MCouCx}) defines the $H$ atom's positions in 
the Riemann $\mathsf{M}_\N$-surface~\ci{MyNDA17,MesResX17,MyAHEP13}; 
 negative eigenmasses correspond to the $H$ anti atom. 
 The centered eigenmasses, $\mathsf{M}_\N^{\rm Re}$, and the total 
widths, $\mathsf{M}_\N^{\rm Im}$, are 
\bea 
\mathsf{M}_\N^{\rm Re} =\pm 2{\rm Re}\{\epsilon_\N^{~}\}\equiv
\pm\sqrt{2\left(|\epsilon_\N^2|+{\rm Re}\{\epsilon_\N^2\}\right)},\\
\label{AtRe}
\mathsf{M}_\N^{\rm Im} =\pm 2{\rm Im}\{\epsilon_\N^{~}\}\equiv
\pm\sqrt{2\left(|\epsilon_\N^2|-{\rm Re}\{\epsilon_\N^2\}\right)}. 
\label{AtIm} \eea
 Note, in case of equal particle masses, $m_1=m_2$ (positronium), 
the imaginary-part mass disappears. 

 It is important to note that (\ref{MCouCx}) has the form of 
the expression for the total energy of two free particles: 
a~proton and an~electron move as free particles inside the $H$ atom. 
 The result is in accordance with the free particle hypothesis for 
bound states.  
 The total kinetic energy of the proton and the electron is 
\be
\mathsf{T}_\N^{\rm H} =|\mathsf{M}_\N^{\rm Re}| -m_p -m_e. 
\ee{AtHkin}

 The spin center-of-gravity energy levels for the $H$ atom calculated 
with the use of (\ref{AtHkin}) are shown in Table~\ref{tab:table1}. 
\begin{table}[tbh]
\caption{
 The $H$ atom energy levels and known results. 
\label{tab:table1}}
\begin{ruledtabular}
\begin{tabular}{crrrr}
\textrm{State}& $\mathsf{T}_\N^{\rm KG}\ \ \ \ \ $ & 
$\mathsf{T}_\N^{\rm SS}\ \ \ \ \ $ &$\mathsf{T}_\N^{\rm QC}
\ \ \ \ \ $ &$\mathsf{T}_\N^{\rm NIST}\ \ \ \ \ $\\
\colrule
$1S$ &-13.60659871&-13.60442520&-13.59810653&-13.59843445\\
$1P$ & -3.40144965& -3.40137418& -3.39956046& -3.39959812\\
$1D$ & -1.51174769& -1.51173516& -1.51091854& -1.51092434\\
$1F$ & -0.85035692& -0.85035328& -0.84989222& -0.84989357\\
$1G$ & -0.54422814& -0.54393117& -0.54393115& -0.54393196\\
$2S$ & -3.40157042& -3.40125344& -3.39956046& -3.39962387\\
$2P$ & -1.51175484& -1.51172801& -1.51091854& -1.51093197\\
$2D$ & -0.85035822& -0.85035199& -0.84989222& -0.84989548\\
$3S$ & -1.51179063& -1.51169223& -1.51091854& -1.51093960\\
$3P$ & -0.85036123& -0.85034897& -0.84989222& -0.84989834
\end{tabular}
\end{ruledtabular}
\end{table}
 Here we compare our calculations $\mathsf{T}_\N^{\rm QC}$ obtained 
from solution of the QC wave equation (\ref{Rel2Equa}) with 
the ones given by the 
static KG ($\mathsf{T}_\N^{\rm KG}$) equation, SS equation 
($\mathsf{T}_\N^{\rm SS}$) and the tabulated NIST 
data ($\mathsf{T}_\N^{\rm NIST}$)~\ci{NIST_ASD}. 

 Our calculations of the $H$ atom spectrum ($\mathsf{T}_\N^{\rm QC}$) 
have precision 3--6 positions and give better description of 
the NIST data in comparison with those obtained from solution of 
the KG ($\mathsf{T}_\N^{\rm KG}$) and SS ($\mathsf{T}_\N^{\rm SS}$) 
equations, which give precision 1--3 positions. 

 In Table~\ref{tab:table2} we represent the relative calculation 
precisions corresponding to the states shown in 
Table~\ref{tab:table1}; 
they are given by $\varepsilon^{\rm eq}
=\left|(\mathsf{T}_\N^{\rm eq}-\mathsf{T}_\N^{\rm NIST})
/\mathsf{T}_\N^{\rm NIST}\right| 100\%$. 
 Also, we bring the imaginary-part masses of the $H$ atom, 
$\mathsf{M}_\N^{\rm Im}$, given by (\ref{AtIm}). 
 If sign is negative, then we come to the relation for the total 
widths, $\GN^\TOT\equiv\GN^{\rm QC}=-2\mathsf{M}_\N^{\rm Im}$. 
\begin{table}[tbh]
\caption{	
 The relative accuracies of the $H$ atom's the energy levels 
and the levels' total width. 
\label{tab:table2}}
\begin{ruledtabular}
\begin{tabular}{ccccc}
\textrm{State}& $\varepsilon_\N^{\rm KG}(\%) $ & 
$\varepsilon_\N^{\rm SS}(\%) $ &$\varepsilon_\N^{\rm QC}(\%)$&
$\mathsf{M}_\N^{\rm Im}$\\
\colrule
$1S$& 6.00$^{-2}$& 4.41$^{-2}$& 2.41$^{-3}$& 3.421587\\
$1P$& 5.45$^{-2}$& 5.22$^{-2}$& 1.11$^{-3}$& 1.710793\\
$1D$& 5.45$^{-2}$& 5.37$^{-2}$& 2.41$^{-3}$& 1.140530\\
$1F$& 5.45$^{-2}$& 5.41$^{-2}$& 3.84$^{-4}$& 0.855397\\
$1G$& 5.45$^{-2}$& 5.42$^{-2}$& 1.59$^{-4}$& 0.684317\\
$2S$& 5.73$^{-2}$& 4.79$^{-2}$& 1.87$^{-4}$& 1.710793\\
$2P$& 5.45$^{-2}$& 5.27$^{-2}$& 8.89$^{-3}$& 1.140530\\
$2D$& 5.54$^{-2}$& 5.37$^{-2}$& 3.84$^{-4}$& 0.855397\\
$3S$& 5.63$^{-2}$& 4.98$^{-2}$& 1.39$^{-3}$& 1.140530\\
$3P$& 5.45$^{-2}$& 5.30$^{-2}$& 7.20$^{-4}$& 0.855397
\end{tabular}
\end{ruledtabular}
\end{table}
 As one can see, our calculation precisions are better of known 
results. 
 More accurate calculations require accounting for the spin 
corrections, i.\,e., fine and hyperfine splittings of the energy 
levels. 
 These corrections were not an objective of our work; 
 they will be considered somewhere else. 

\section{Discussion and Conclusion}
 The potential approach, in spite of nonrelativistic 
phenomenological nature the potential, is used with success 
to describe the $H$ atom spectrum. 
 We have considered the $H$ atom as a relativistic two-body 
problem for the static Lorentz-scalar Coulomb potential and 
took into account the proton structure, using the dipole 
form factor, and motion effects. 
 We have derived relativistic two-body wave equation and found, 
from its solution, the complex-mass expression for the $H$ 
atom's eigenstates. 
 In the framework of developed approach we have calculated 
the spin center-of-gravity energy levels for the $H$ atom and
compared them with the ones obtained from solution of 
the Klein-Gordon, spinless Salpeter equations and tabulated 
NIST data. 

 In our consideration, we have came to free particle hypothesis 
and those relevant to particles subject to non-trivial potentials 
like the modified Coulomb potential. 
 The $H$ atom can be used as a tool for testing any relativistic 
two-body theory, because latest measurements for transition 
frequencies have been determined with a highest 
precision\ci{NIST_ASD,MohrTayl,CharlTab93}.

 Description of the $H$ atom with the use of spinless equations 
is not exact. 
 There are various relativistic effects such as fine and hyperfine 
splitting of the energy levels that should be taken into account 
in more accurate analysis. 
 The corrections are applied using perturbation theory. 
 Hyperfine splitting couples the spins of proton and electron, 
and in the ground state, they combine in the singlet state. 
 A slightly higher energy level occurs when they are in a spin-one 
triplet state. 
 Transitions between these states radiate very low energy photons 
with a~wave length of 21~cm. 
 This is the source of the 21 centimeter line or ``hydrogen line 
radiation'' that is of great importance in cosmology. 
 It has been used to analyze the spiral arms of the galaxy, 
and can shed light on the so called dark ages that the universe 
went through. 
 Errors in the ground state energy of hydrogen that are 
$\sim 10^{-6}$ the energy itself can be of critical importance. 
 This effect is a factor $m_e/m_p\propto 10^{-3}$ smaller still 
than the fine structure corrections, making the associated energy 
changes about two orders of magnitude smaller. 

\newpage
\bibliography{RelHpre}

\end{document}